\pdfoutput=1

\documentclass[twocolumn]{jpsj3}
\usepackage{bm}
\usepackage{tabularx}
\usepackage{ascmac}
\usepackage{amsmath}
\usepackage{wrapfig}
\usepackage{graphicx}
\usepackage{float}
\usepackage[FIGBOTCAP]{subfigure}
\usepackage{color}

\bmdefine{\boldi}{i}
\bmdefine{\boldj}{j}
\bmdefine{\boldr}{r}
\bmdefine{\boldS}{S}
\bmdefine{\boldq}{q}
\bmdefine{\boldk}{k}
\bmdefine{\boldzero}{0}
\bmdefine{\bolddelta}{\delta}

\title{Mechanism for a Chemical Potential of Nonequilibrium Magnons
  in Parametric Parallel Pumping }

\author{Naoya Arakawa\thanks{E-mail address: naoya.arakawa@sci.toho-u.ac.jp}}
\inst{Department of Physics, Toho University, 
Funabashi, Chiba, 274-8510, Japan}

\abst{
  We demonstrate how a magnon chemical potential is generated 
  in parametric parallel pumping.
  We study how a time-periodic magnetic field of this pumping 
  affects magnon properties of a ferrimagnet in a nonequilibrium steady state. 
  We show that
  the magnon distribution function of our nonequilibrium steady state
  becomes the Bose distribution function with
  $\mu=\omega_{\textrm{p}}/2$, where $\mu$ is the magnon chemical potential
  and $\omega_{\textrm{p}}$ is the pumping frequency. 
  This result is distinct from
  the absence of the magnon chemical potential in the standard theory
  and can qualitatively explain its generation in experiments.  
  We believe our result is a first theoretical demonstration of
  the generation of the magnon chemical potential
  in the parametric parallel pumping,
  providing an important step towards a thorough understanding of properties of 
  nonequilibrium magnons.  
}

\begin{document}
\maketitle

\section{Introduction}

A magnon chemical potential
is a key parameter in magnon Bose-Einstein condensation (BEC)
and transport phenomena. 
Magnons are bosonic quasiparticles that describe the collective motions of a magnet.
To realize the magnon BEC~\cite{MagBEC,MagBEC-Nature},
the magnon chemical potential $\mu$
should satisfy $\epsilon_{0}-\mu=0$,
where $\epsilon_{0}$ denotes the lowest energy of magnon bands. 
Since $\epsilon_{0}$ can be a nonzero positive value,
tuning the value of $\mu$
is necessary for the magnon BEC.
Then
$\mu$ plays an essential role in transport phenomena 
for a multilayer including a magnet~\cite{NatPhys-mu,PRL-mu,PRB-mu,Science-mu,NatCom-mu}. 
For example,
a change of $\mu$ near the interface 
needs to be taken into account
in estimating spin transport in the spin Seebeck effect for a bilayer
of Pt and yttrium iron garnet (YIG), a ferrimagnet~\cite{PRB-mu}. 

\begin{figure}
  \begin{center}
    \includegraphics[width=52mm]{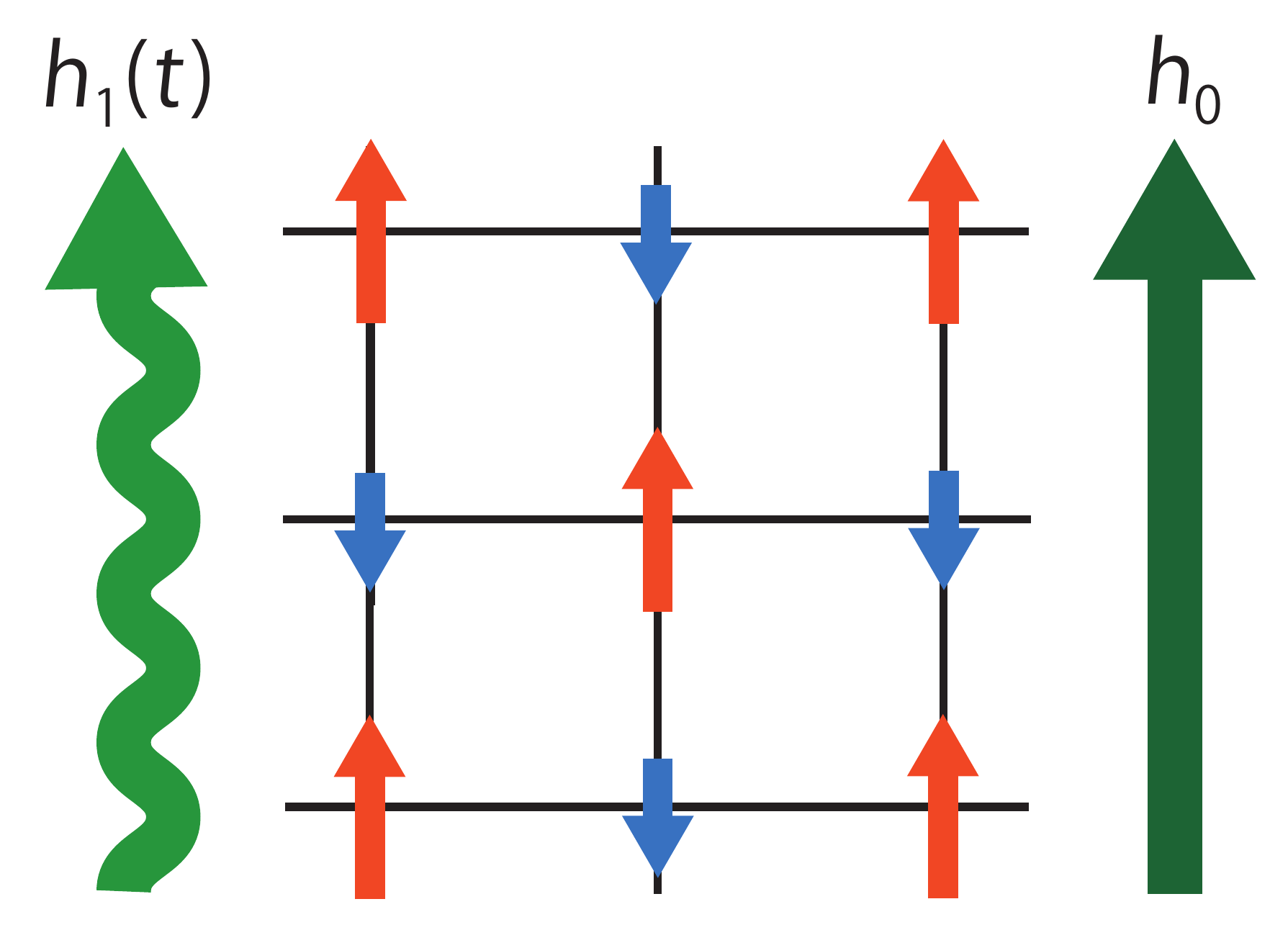}
  \end{center}
\caption{\label{fig1} 
  (Color online) Setup of the parametric parallel pumping of a ferrimagnet.
  As a simple case, a two-sublattice ferrimagnet is considered.
  The time-periodic magnetic field $h_{1}(t)$ (a green wavy line)
  is used to generate $\mu$; 
  the time-independent magnetic field $h_{0}$ (a green straight line)
  is used to align the magnetization direction along it.  
}
\end{figure}

Despite progress in understanding $\mu$, 
there exists a gap between experiment and theory. 
From an experimental point of view,
$\mu$ can be finite
by using parametric parallel pumping~\cite{MagBEC-Nature,Estimate-MagChem}. 
This method~\cite{PP1,PP2,PP3,PP4} uses
two different magnetic fields parallel to each other (Fig. \ref{fig1}): 
a time-independent one $h_{0}$ and a time-periodic one $h_{1}(t)$
with a period of $T=2\pi/\omega_{\textrm{p}}$. 
In this pumping the system of magnons is nonequilibrium. 
After a certain period of time
the system can achieve a quasiequilibrium state in which
the magnon distribution function can be approximated
by the Bose distribution function with finite $\mu$~\cite{MagBEC-Nature,Estimate-MagChem}.
However, from a theoretical point of view, 
it remains unclear how $\mu$ can be generated under $h_{1}(t)$.
In the standard theory~\cite{S-theory1,S-theory2,S-theory3,PRB-Stheory1,Inv-Stheory},
which is sometimes called the $S$-theory, 
$h_{1}(t)$ is treated as a classical field in the form $h\cos(\omega_{\textrm{p}}t)$, 
and its effect is described by the Hamiltonian 
$H_{\textrm{pump}}(t)
=g\mu_{\textrm{B}}h_{1}(t)\sum\textstyle_{\boldj}S^{z}_{\boldj}$, where 
$g$ is the $g$ factor, $\mu_{\textrm{B}}$ is the Bohr magnetron,
and $S^{z}_{\boldj}$ is the $z$-component of the spin operator at cite $\boldj$. 
$H_{\textrm{pump}}(t)$ is then rewritten
as the magnons-pair creation and annihilation terms
by using the Holstein-Primakoff transformation~\cite{HP} and several approximations.  
Since such terms violate the magnon-number conservation,
this theory leads to $\mu=0$~\cite{S-theory2,Inv-Stheory}.
(Note that a chemical potential of bosons or fermions becomes zero
when the number is not conserved~\cite{Kubo-text2}.)
This theoretical result (i.e., $\mu=0$) implies that
in the case of nonzero $\epsilon_{0}$  
it is impossible to realize the BEC of magnons.
Thus there is the gap between experiment and theory,
and its existence may imply that
something is missing in the standard theory.

In this paper
we present a new theory of the parametric parallel pumping,
and we demonstrate a mechanism by which the magnon chemical potential is generated. 
We first introduce a model Hamiltonian for a ferrimagnet in the parametric parallel pumping,
and then derive the master equation of the reduced density matrix of magnons. 
We show that
the nonequilibrium steady state is achieved
due to the detailed balance between the magnons-pair creation and annihilation. 
Most importantly, 
the magnon distribution function of this steady state 
is the Bose distribution function with $\mu=\omega_{p}/2$.
This result is, to the best of author's knowledge, 
a first theoretical demonstration of generation of the magnon chemical potential
in the parametric parallel pumping.

The rest of this paper is organized as follows.
In Sect. 2 we derive the model Hamiltonian for a two-sublattice ferrimagnet
in the parametric parallel pumping.
Our Hamiltonian consists of the magnon Hamiltonian of the ferrimagnet,
the magnon-photon coupling Hamiltonian due to the time-dependent magnetic field,
and the photon Hamiltonian.
In contrast to the standard theory~\cite{S-theory1,S-theory2,S-theory3,PRB-Stheory1,Inv-Stheory}, 
the time-periodic magnetic field is treated as a quantized field in our theory. 
We also argue that
our two-sublattice ferrimagnet can be regarded as a minimal model
for describing magnon properties of YIG at room temperature.
In Sect. 3
we derive the equation of motion of the reduced density matrix of magnons
and write it in the form of the master equation.
In this derivation
we treat photons as a Markovian bath for magnons
and assume that the magnon-photon coupling is weak enough
to treat its Hamiltonian as perturbation.
Such a treatment of photons may be appropriate for YIG,
in which the magnon lifetime is sufficiently long~\cite{Bauer-photon}.
In Sect. 4
we study a steady-state solution to the master equation,
and we show the magnon properties in the nonequilibrium steady state
for the parametric parallel pumping.
In Sect. 5
we compare our result with the experimental results, and 
we discuss the differences between our theory and the standard theory
and the implications of our theory.
In Sect. 6
we summarize the achievements of this paper. 
Throughout this paper we take $\hbar=1$.

\section{Model Hamiltonian}

Our model Hamiltonian is 
\begin{align}
  H=H_{\textrm{s}}+H_{\textrm{sb}}(t)+H_{\textrm{b}},\label{eq:Htot}
\end{align}
where
$H_{\textrm{s}}$, $H_{\textrm{sb}}(t)$, and $H_{\textrm{b}}$
are the system Hamiltonian,
the system-bath coupling Hamiltonian, 
and the bath Hamiltonian, respectively.
As we will explain below,
$H_{\textrm{s}}$, $H_{\textrm{sb}}(t)$, and $H_{\textrm{b}}$ are 
given by the magnon Hamiltonian for a ferrimagnet [Eq. (\ref{eq:Hs})], 
the magnon-photon coupling Hamiltonian [Eq. (\ref{eq:Hsb})],
and the photon Hamiltonian [Eq. (\ref{eq:Hb})], respectively.

We first derive $H_{\textrm{s}}$.
Since a two-sublattice Heisenberg ferrimagnet~\cite{Nakamura,NA-Lett} 
is a minimal model for a ferrimagnet, 
we consider the following Hamiltonian:
\begin{align}
  H_{\textrm{s}}
  =2J\sum\limits_{\langle\boldi,\boldj\rangle}
  \boldS_{\boldi}\cdot\boldS_{\boldj}
  +g\mu_{\textrm{B}}h_{0}\sum\limits_{\boldj}S_{\boldj}^{z},\label{eq:H_spin}
\end{align}
where the sum $\sum_{\langle\boldi,\boldj\rangle}$ is restricted to
nearest-neighbor sites for $\boldi\in A$, $\boldj\in B$.
For simplicity we suppose that
the numbers of the $A$ sublattice and the $B$ sublattice are $N/2$. 
In Eq. (\ref{eq:H_spin})
the first term corresponds to the Heisenberg Hamiltonian of a two-sublattice ferrimagnet,
and the second term corresponds to the Zeeman coupling Hamiltonian
due to the time-independent magnetic field $h_{0}$. 
The spin Hamiltonian of Eq. (\ref{eq:H_spin}) can be rewritten as the magnon Hamiltonian
by using the following Holstein-Primakoff transformation~\cite{Nakamura,NA-Lett}: 
\begin{eqnarray}
  \hspace{-8pt}
S_{\boldi}^{z}=S_{A}-a_{\boldi}^{\dagger}a_{\boldi},
S_{\boldi}^{-}=a_{\boldi}^{\dagger}
\sqrt{2S_{A}-a_{\boldi}^{\dagger}a_{\boldi}},
S_{\boldi}^{+}=(S_{\boldi}^{-})^{\dagger},\ \label{eq:HP-A}\\
  \hspace{-8pt}
S_{\boldj}^{z}=-S_{B}+b_{\boldj}^{\dagger}b_{\boldj},  
S_{\boldj}^{+}=b_{\boldj}^{\dagger}
\sqrt{2S_{B}-b_{\boldj}^{\dagger}b_{\boldj}},
S_{\boldj}^{-}=(S_{\boldj}^{+})^{\dagger},\ \label{eq:HP-B}
\end{eqnarray}
where
$a_{\boldi}$ and $a_{\boldi}^{\dagger}$ are
the annihilation and creation operators of a magnon for the $A$ sublattice, 
and $b_{\boldj}$ and $b_{\boldj}^{\dagger}$ are those for the $B$ sublattice.
Although substitution of Eqs. (\ref{eq:HP-A}) and (\ref{eq:HP-B})
into the first term of Eq. (\ref{eq:H_spin}) leads to not only the kinetic energy terms
but also the interaction terms of magnons~\cite{Nakamura,NA-Lett},
we consider only the kinetic energy terms for simplicity. 
After some algebra~\cite{Nakamura,NA-Lett,NA-AF},
we can rewrite Eq. (\ref{eq:H_spin}) as
\begin{align}
  H_{\textrm{s}}
  =&2\sum\limits_{\boldq}J(\boldzero)
  (S_{B}a_{\boldq}^{\dagger}a_{\boldq}+S_{A}b_{\boldq}^{\dagger}b_{\boldq})\notag\\
  &+2\sum\limits_{\boldq}J(\boldq)\sqrt{S_{A}S_{B}}
  (a_{\boldq}b_{\boldq}+a_{\boldq}^{\dagger}b_{\boldq}^{\dagger})\notag\\
  &-h_{0}M
  -g\mu_{\textrm{B}}h_{0}\sum\limits_{\boldq}
  (a_{\boldq}^{\dagger}a_{\boldq}-b_{\boldq}^{\dagger}b_{\boldq}),\label{eq:Hs-sub}
\end{align}
where 
\begin{align}
  a_{\boldi}
  &=\sqrt{\tfrac{2}{N}}\sum_{\boldq}e^{i\boldq\cdot\boldi}a_{\boldq},\\ 
  b_{\boldj}^{\dagger}
  &=\sqrt{\tfrac{2}{N}}\sum_{\boldq}e^{i\boldq\cdot\boldj}b_{\boldq}^{\dagger},\\ 
  J(\boldq)
  &=\sum_{\bolddelta}Je^{i\boldq\cdot\bolddelta},
\end{align}
with $\bolddelta$ being a vector to nearest neighbors;
and $M$ is the magnetization without magnons,
\begin{align}
  M=&(-g\mu_{\textrm{B}})\frac{N}{2}S_{A}+(-g\mu_{\textrm{B}})\frac{N}{2}(-S_{B})\notag\\
  =&g\mu_{\textrm{B}}\frac{N}{2}(S_{B}-S_{A}).\label{eq:Magnetization}
\end{align}
In Eq. (\ref{eq:Hs-sub}) we have neglected 
the constant terms arising from the Heisenberg interaction.
In the following analyses
we also neglect the term of $-h_{0}M$ in Eq. (\ref{eq:Hs-sub})
because its role is just to make the directions of the time-independent magnetic field
and the magnetization parallel. 
By using the Bogoliubov transformation,
\begin{align}
  a_{\boldq}&=\cosh\theta_{\boldq}\alpha_{\boldq}
  -\sinh\theta_{\boldq}\beta_{\boldq}^{\dagger},\label{eq:Bogo1}\\ 
  b_{\boldq}^{\dagger}&=-\sinh\theta_{\boldq}\alpha_{\boldq}
  +\cosh\theta_{\boldq}\beta_{\boldq}^{\dagger},\label{eq:Bogo2}
\end{align}
where
\begin{align}
  \tanh2\theta_{\boldq}=\frac{2\sqrt{S_{A}S_{B}}J(\boldq)}{(S_{A}+S_{B})J(\boldzero)},
\end{align}
we can diagonalize Eq. (\ref{eq:Hs-sub})
as follows~\cite{Nakamura,NA-Lett,NA-AF}:
\begin{align}
  H_{\textrm{s}}
  =
  \sum\limits_{\boldq}\tilde{\epsilon}_{\alpha}(\boldq)\alpha_{\boldq}^{\dagger}\alpha_{\boldq}
  +\sum\limits_{\boldq}\tilde{\epsilon}_{\beta}(\boldq)\beta_{\boldq}^{\dagger}\beta_{\boldq},\label{eq:Hs}
\end{align}
where
\begin{align}
  \tilde{\epsilon}_{\alpha}(\boldq)
  &=\epsilon_{\alpha}(\boldq)-g\mu_{\textrm{B}}h_{0}\notag\\
  &=(S_{B}-S_{A})J(\boldzero)+\Delta\epsilon(\boldq)-g\mu_{\textrm{B}}h_{0},\label{eq:E_alpha}\\
  \tilde{\epsilon}_{\beta}(\boldq)
  &=\epsilon_{\beta}(\boldq)+g\mu_{\textrm{B}}h_{0}\notag\\
  &=(S_{A}-S_{B})J(\boldzero)+\Delta\epsilon(\boldq)+g\mu_{\textrm{B}}h_{0},\label{eq:E_beta}
\end{align}
and 
\begin{align}
  \Delta\epsilon(\boldq)
  =\sqrt{(S_{A}+S_{B})^{2}J(\boldzero)^{2}-4S_{A}S_{B}J(\boldq)^{2}}.\label{eq:DelE}
\end{align}
As we will show in Appendix A,
the $h_{0}$ makes 
the lowest energy of the magnon bands nonzero. 

Before the derivation of $H_{\textrm{sb}}(t)$,
we argue the validity of the above model in describing
magnon properties of YIG at room temperature.
Although YIG is a ferrimagnet,
its magnon properties have been often discussed
by using 
magnons of a ferromagnet with no sublattice.
However,
a theoretical study~\cite{Bauer-Ferri} using a ferrimagnetic Heisenberg model for YIG 
has shown that
it is necessary to take account of
not only the lowest-energy branch of magnon bands,
which can be approximately described by magnons of the ferromagnet,
but also
the second-lowest-energy branch
for describing magnon properties of YIG at room temperature.
Since the magnon spectrum obtained in that study~\cite{Bauer-Ferri}
agrees very well with the results of neutron scattering experiments~\cite{Neutron-YIG},
the above result indicates that
in order to describe magnon properties of YIG at room temperature,
one needs to consider, at least, two magnon bands.
Note that 
in that theoretical study~\cite{Bauer-Ferri}
the magnetic anisotropy and dipolar interaction are neglected
because they are much smaller than the Heisenberg exchange interactions.
Actually, 
another theoretical study~\cite{npj-YIG} has shown that
the effects of the magnetic anisotropy terms on the magnon spectrum of YIG
are vanishingly small.
Then first-principles calculations~\cite{YIG-1stPrinciple} of YIG
have shown that
the largest term of the Heisenberg exchange interactions
is the antiferromagnetic nearest-neighbor Heisenberg exchange interaction
between Fe$^{\textrm{O}}$ and Fe$^{\textrm{T}}$ ions,
which are Fe ions surrounded by an octahedron and a tetrahedron of O ions, respectively,
and the other terms are at least an order of magnitude smaller. 
Since these facts can be taken into account in our two-sublattice ferrimagnet,
we believe that
our model can be regarded as a minimal model for describing magnon properties of YIG
at room temperature. 

We then derive $H_{\textrm{sb}}(t)$ in a way different from that of the standard theory. 
We suppose that
the main effect of a time-periodic magnetic field $h_{\boldr}(t)$
can be described by 
\begin{align}
  H_{\textrm{sb}}(t)=g\mu_{\textrm{B}}\sum\limits_{\boldr}h_{\boldr}(t)S^{z}_{\boldr}.\label{eq:Hsb-pre} 
\end{align}
In contrast to the standard theory~\cite{S-theory1,S-theory2,S-theory3,PRB-Stheory1,Inv-Stheory}, 
we treat the time-periodic magnetic field as a quantized field.
(This is because 
its time dependence can be appropriately described only for a quantum theory;
if the time-periodic magnetic field is treated in a classical theory, 
an approximation whose validity is uncertain
is used~\cite{Inv-Stheory}.) 
First, 
the quantized magnetic field is expressed in the form~\cite{Schiff}
\begin{align}
  \hspace{-10pt}
  h_{\boldr}(t)
  =\sum\limits_{\boldk,\lambda}
  [C_{\boldk\lambda}e^{i(\boldk\cdot \boldr-\omega_{\boldk}t)}c_{\boldk\lambda}
  +C_{\boldk\lambda}^{\ast}e^{-i(\boldk\cdot \boldr-\omega_{\boldk}t)}c_{\boldk\lambda}^{\dagger}],\label{eq:mag-quant} 
\end{align}
where 
$c_{\boldk\lambda}$ and $c_{\boldk\lambda}^{\dagger}$
are the annihilation and creation operators of a photon
for $\omega_{\boldk}=c|\boldk|$ with the mode index $\lambda$.
(We have not explicitly expressed the coefficient $C_{\boldk\lambda}$  
because its detail is irrelevant to the steady-state properties.) 
Since $\omega_{\boldk}$ is chosen to be $\omega_{\boldk}=\omega_{\textrm{p}}$ in the parametric pumping,
we replace $e^{\mp i\omega_{\boldk}t}$ in Eq. (\ref{eq:mag-quant})
by $e^{\mp i\omega_{\textrm{p}}t}\delta(\omega_{\boldk}-\omega_{\textrm{p}})$. 
Then we express $S^{z}_{\boldr}$ in terms of the magnon operators
by using Eqs. (\ref{eq:HP-A}) and (\ref{eq:HP-B}). 
Combining these results with Eq. (\ref{eq:Hsb-pre})
and using the Fourier transformations of the magnon operators,
we obtain
\begin{align}
  H_{\textrm{sb}}(t)=
  &\sum\limits_{\boldq,\boldq^{\prime},\lambda}
  \tilde{C}_{\boldq-\boldq^{\prime}\lambda}
  e^{-i\omega_{\textrm{p}}t}c_{\boldq-\boldq^{\prime}\lambda}
  (b_{\boldq^{\prime}}^{\dagger}b_{\boldq}-a_{\boldq}^{\dagger}a_{\boldq^{\prime}})+(\textrm{H.c.}),\label{eq:Hsb-ab}
\end{align}
where $\tilde{C}_{\boldk\lambda}=g\mu_{\textrm{B}}C_{\boldk\lambda}\delta(\omega_{\boldk}-\omega_{\textrm{p}})$.   
We can also represent Eq. (\ref{eq:Hsb-ab})
in terms of the magnon-band operators 
by using the Bogoliubov transformation and
retaining only the relevant terms (see Appendix B): 
\begin{align}
  H_{\textrm{sb}}(t)
  =e^{-i\omega_{\textrm{p}}t}H_{\textrm{sb}}^{(\textrm{abs})}
  +e^{i\omega_{\textrm{p}}t}H_{\textrm{sb}}^{(\textrm{emi})},\label{eq:Hsb}
\end{align}
where 
\begin{align}
  &H_{\textrm{sb}}^{(\textrm{abs})}
  =\sum\limits_{\boldq,\boldq^{\prime},\lambda}
  \tilde{C}_{\boldq-\boldq^{\prime}\lambda}
  c_{\boldq-\boldq^{\prime}\lambda}
  B_{\boldq\boldq^{\prime}}
  \alpha_{\boldq}^{\dagger}\beta_{\boldq^{\prime}}^{\dagger},\label{eq:Hsb^abs}\\
  &H_{\textrm{sb}}^{(\textrm{emi})}
  =\sum\limits_{\boldq,\boldq^{\prime},\lambda}
  \tilde{C}_{\boldq-\boldq^{\prime}\lambda}^{\ast}
  c_{\boldq-\boldq^{\prime}\lambda}^{\dagger}
  B_{\boldq\boldq^{\prime}}
  \alpha_{\boldq}\beta_{\boldq^{\prime}},\label{eq:Hsb^emi}  
\end{align}
and
\begin{align}
  B_{\boldq\boldq^{\prime}}=\cosh\theta_{\boldq}\sinh\theta_{\boldq^{\prime}}
  -\sinh\theta_{\boldq}\cosh\theta_{\boldq^{\prime}}.\label{eq:Bqq'}
\end{align}
Thus the main effect of $h_{\boldr}(t)$ is to create and annihilate a pair of magnons
in different bands. 
Although the terms of Eqs. (\ref{eq:Hsb^abs}) and (\ref{eq:Hsb^emi}) violate
magnon-number conservation in general,
the rates of the pair creation and the pair annihilation
satisfy the detailed balance in our nonequilibrium steady state;
as a result,
the effects of the $H_{\textrm{sb}}^{(\textrm{abs})}$ and $H_{\textrm{sb}}^{(\textrm{emi})}$
can be reduced to a nonzero chemical potential of nonequilibrium magnons
(see Sect. 4). 

In addition to the magnon-photon Hamiltonian,
we consider the photon Hamiltonian~\cite{Schiff}.
It is
\begin{align}
  H_{\textrm{b}}=\sum\limits_{\boldk,\lambda}
  \omega_{\boldk}c_{\boldk\lambda}^{\dagger}c_{\boldk\lambda}.\label{eq:Hb}
\end{align}

\section{Master equation}

We derive the equation of motion of the reduced density matrix of magnons for our system,
and we express it in the form of the master equation.
The following derivation is an extension of
that for an electron system~\cite{Kubo-text,Master1,Master2,Master3}. 

In the following analyses
we use several approximations.
To take account of a finite lifetime of magnons or photons,
we introduce the lifetime of magnons, $\tau_{\textrm{m}}$,
and the lifetime of photons, $\tau_{\textrm{p}}$,
in a phenomenological way,
such as the relaxation-time approximation for an electron system~\cite{Ashcroft}.
(Such finite lifetimes are induced, for example, by the scattering of impurities.) 
We assume that $\tau_{\textrm{m}}\gg \tau_{\textrm{p}}$,
which is valid for YIG~\cite{Bauer-photon}.
Then  
we suppose that
the $H_{\textrm{sb}}(t)$ is weak enough to treat it as perturbation. 
[More precisely,
it is so weak that
$\tau_{\textrm{r}} \gg \tau_{\textrm{p}}$, where $\tau_{\textrm{r}}$ is 
the relaxation time of magnons
due to the second-order perturbation of the $H_{\textrm{sb}}(t)$
and characterizes
a time evolution of the reduced density matrix of magnons.] 
We also suppose that
$\tau_{\textrm{r}} < \tau_{\textrm{m}}$, which is valid for YIG~\cite{Estimate-MagChem}.
Under those conditions,
photons can be treated as a Markovian bath for magnons~\cite{Bauer-photon}, 
and the $H_{\textrm{sb}}(t)$ can be regarded as the system-bath coupling Hamiltonian.
Since the bath degrees of freedom
can be traced over~\cite{Kubo-text,Master1,Master2,Master3}
in the equation of motion of the density matrix for $H$, 
dynamics of nonequilibrium magnons for our system 
can be described by
the equation of motion of the reduced density matrix of magnons
which are weakly coupled to a Markovian bath of photons.

We can derive the equation of motion
of the reduced density matrix of the magnons as follows. 
The dynamics for $H$ of Eq. (\ref{eq:Htot})
can be described by the Liouville equation,
\begin{align}
  \frac{d \rho(t)}{d t}
  =\frac{1}{i}[H,\rho(t)],\label{eq:Liouville}
\end{align}
where
$\rho(t)$ is the density matrix for $H$.
To describe magnon dynamics,
we rewrite Eq. (\ref{eq:Liouville})
as the equation of motion of the reduced density matrix of magnons,
\begin{align}
  \rho_{\textrm{s}}(t)=\textrm{tr}_{\textrm{b}}\rho(t),\label{eq:rho_s}
\end{align}
where $\textrm{tr}_{\textrm{b}}$ denotes a trace over the bath variables.
This can be done in a manner similar to
the derivation for an electron system~\cite{Kubo-text,Master1,Master2,Master3}. 
Since the details of that derivation have been described
in several textbooks (e.g., Ref. \cite{Kubo-text}),
we quote an expression here:
\begin{align}
  \frac{d \rho_{\textrm{s}}^{(\textrm{I})}(t)}{d t} 
  = -\textrm{tr}_{\textrm{b}}[H_{\textrm{sb}}^{(\textrm{I})}(t),\int_{0}^{t}d\tau
    [H_{\textrm{sb}}^{(\textrm{I})}(\tau),\rho_{\textrm{b}}\rho_{\textrm{s}}^{(\textrm{I})}(t)]],\label{eq:approx}
\end{align}
where the operators in the interaction picture,
$\rho_{\textrm{s}}^{(\textrm{I})}(t)$ and $H_{\textrm{sb}}^{(\textrm{I})}(t)$,
are defined as
\begin{align}
  \rho_{\textrm{s}}^{(\textrm{I})}(t)
  &=e^{iH_{\textrm{s}}t}\rho_{\textrm{s}}(t)e^{-iH_{\textrm{s}}t},\label{eq:rho_s^I}\\
  H_{\textrm{sb}}^{(\textrm{I})}(t)
  &=e^{i(H_{\textrm{s}}+H_{\textrm{b}})t}H_{\textrm{sb}}(t)e^{-i(H_{\textrm{s}}+H_{\textrm{b}})t},\label{eq:H_sb^I}
\end{align}
and $\rho_{\textrm{b}}$ is the density matrix of photons. 
[For the derivation of Eq. (\ref{eq:approx}), see Appendix C with Appendix D.] 

To proceed further
we rewrite Eq. (\ref{eq:approx}) 
as the equation for the diagonal elements of $\rho_{\textrm{s}}(t)$
for the eigenstates of $H_{\textrm{s}}$. 
Let us introduce $|m\rangle$, 
an eigenvector of $H_{\textrm{s}}$: 
$H_{\textrm{s}}|m\rangle=E_{m}|m\rangle$.
This $|m\rangle$ also satisfies
$N_{\textrm{s}}|m\rangle=N_{m}|m\rangle$,
where $N_{\textrm{s}}$ is the operator of the total number of magnons
and $N_{m}$ is its value for $|m\rangle$.
This is because $H_{\textrm{s}}$ of Eq. (\ref{eq:Hs}) does not violate the number conservation.
(This property may hold approximately
even in the presence of interactions of magnons
for the temperatures lower than the Curie temperature 
because for such temperatures the number-nonconserving terms of the interactions
are negligible compared with the number-conserving terms~\cite{PRB-mu}.)
By using $|m\rangle$, 
we define the diagonal elements of $\rho_{\textrm{s}}(t)$ as 
$p_{m}(t)=\langle m|\rho_{\textrm{s}}(t)|m\rangle$,
where $p_{m}(t)$ represents the occupation probability of magnons.
In addition, 
to trace over the bath variables in Eq. (\ref{eq:approx}),
we introduce $|p\rangle$, an eigenvector of $H_{\textrm{b}}$:
$H_{\textrm{b}}|p\rangle=E_{p}|p\rangle$.
Since $\frac{d }{dt}p_{m}(t)
=\langle m|\frac{d }{dt}\rho_{\textrm{s}}^{(\textrm{I})}(t)|m\rangle$,
Eq. (\ref{eq:approx}) can be rewritten as 
\begin{align}
  \frac{d p_{m}(t)}{dt}
  =-\sum\limits_{m^{\prime}}R_{mm^{\prime}}(t)p_{m}(t)
  +\sum\limits_{m^{\prime}}R_{m^{\prime}m}(t)p_{m^{\prime}}(t),\label{eq:Master-tdep-pre}
\end{align}
where
\begin{align}
  &R_{mm^{\prime}}(t)
  =
  \int_{0}^{t}d\tau \sum\limits_{p,p^{\prime}}p_{p}
  \{
  \langle m|\langle p|H_{\textrm{sb}}(t)|p^{\prime}\rangle|m^{\prime}\rangle\notag\\
  &\times
  \langle m^{\prime}|\langle p^{\prime}| H_{\textrm{sb}}(\tau)|p\rangle|m\rangle
  e^{i\Delta E(t-\tau)}+(\textrm{H.c.})
  \},\label{eq:R-pre}
\end{align}
with 
$p_{p}=\langle p|\rho_{\textrm{b}}|p\rangle$ and 
$\Delta E=E_{m}+E_{p}-E_{m^{\prime}}-E_{p^{\prime}}$
(for the details see Appendix E).
Here the $p_{p}$, the occupation probability of photons, is given by
\begin{align}
  p_{p}=\frac{e^{-\beta E_{p}}}{\sum\limits_{p^{\prime\prime}}e^{-\beta E_{p^{\prime\prime}}}},\label{eq:p_p}   
\end{align}
where $\beta=(k_{\textrm{B}}T)^{-1}$.
(Note that the $p_{p}$ can be approximated by the equilibrium occupation probability
because the photons can be treated as a bath for magnons.)
The time integration in Eq. (\ref{eq:R-pre}) can be performed 
with the use of Eq. (\ref{eq:Hsb}); 
the result is
\begin{align}
  &R_{mm^{\prime}}(t)
  =\sum\limits_{p,p^{\prime}}
  |\langle m^{\prime}|\langle p^{\prime}|H_{\textrm{sb}}^{(\textrm{emi})}|p\rangle|m\rangle|^{2}
  p_{p}\frac{2\sin\Delta E_{-}t}{\Delta E_{-}}
  \notag\\
  &+\sum\limits_{p,p^{\prime}}
  |\langle m^{\prime}|\langle p^{\prime}|H_{\textrm{sb}}^{(\textrm{abs})}|p\rangle|m\rangle|^{2}
  p_{p}\frac{2\sin\Delta E_{+}t}{\Delta E_{+}},\label{eq:R}
\end{align}
where $\Delta E_{\mp}=\Delta E\mp\omega_{\textrm{p}}$ (see Appendix F).
Since $R_{mm^{\prime}}(t)$ is
the transition rate of the magnon system from $|m\rangle$ to $|m^{\prime}\rangle$,
Eq. (\ref{eq:Master-tdep-pre}) is the master equation
for the magnon system that is weakly coupled to the Markovian bath.

We remark on Eq. (\ref{eq:Master-tdep-pre}). 
The first term on its right-hand side 
denotes the contribution due to the transitions from $|m\rangle$ to $|m^{\prime}\rangle$,
whereas 
the second term 
denotes the contribution due to the transitions from $|m^{\prime}\rangle$ to $|m\rangle$.
Since these contributions are not balanced in general,
the expectation value of the magnon number,
$\langle N_{\textrm{s}}\rangle=\sum_{m}N_{m}p_{m}(t)$,
should depend on time except the steady-state case.
In such time-dependent cases,
the magnon number is not conserved,
and thus the magnon chemical potential should be zero.
However,
the magnon chemical potential could be finite
in the steady-state case
because the $\langle N_{\textrm{s}}\rangle$ becomes independent of time.
We will demonstrate this property in the next section. 

\section{Steady-state solution}

We now study the steady-state solution
to Eq. (\ref{eq:Master-tdep-pre}). 
Since we focus on the nonequilibrium steady state
that is achieved after a long time evolution
under the time-periodic magnetic field,
we replace the factors $\frac{2\sin\Delta E_{\mp}t}{\Delta E_{\mp}}$ in Eq. (\ref{eq:R}) 
by $2\pi\delta(\Delta E_{\mp})$;
this replacement is valid for large $t$.
Thus Eq. (\ref{eq:R}) becomes
\begin{align}
  R_{mm^{\prime}}(t)\sim \bar{R}_{mm^{\prime}}^{(-)}+\bar{R}_{mm^{\prime}}^{(+)},\label{eq:R-bar}
\end{align}
where  
\begin{align}
  &\bar{R}_{mm^{\prime}}^{(-)}
  = 2\pi \sum\limits_{p,p^{\prime}}
  |\langle m^{\prime}|\langle p^{\prime}|H_{\textrm{sb}}^{(\textrm{emi})}|p\rangle|m\rangle|^{2}
  p_{p}\delta(\Delta E_{-}),\label{eq:R-bar^-}\\
  &\bar{R}_{mm^{\prime}}^{(+)}
  = 2\pi \sum\limits_{p,p^{\prime}}
  |\langle m^{\prime}|\langle p^{\prime}|H_{\textrm{sb}}^{(\textrm{abs})}|p\rangle|m\rangle|^{2}
  p_{p}\delta(\Delta E_{+}).\label{eq:R-bar^+}
\end{align}
$\bar{R}_{mm^{\prime}}^{(-)}$ and $\bar{R}_{mm^{\prime}}^{(+)}$ correspond to 
the transition rates given by Fermi's golden rule.
Since the steady-state solution
to Eq. (\ref{eq:Master-tdep-pre}), $\bar{p}_{m}$,
satisfies $\frac{d}{dt}\bar{p}_{m}=0$, 
$\bar{p}_{m}$ is determined by
\begin{align}
  \hspace{-10pt}
  0=\sum\limits_{m^{\prime}}
  \{[\bar{R}_{mm^{\prime}}^{(-)}+\bar{R}_{mm^{\prime}}^{(+)}]\bar{p}_{m}
    -[\bar{R}_{m^{\prime}m}^{(-)}+\bar{R}_{m^{\prime}m}^{(+)}]\bar{p}_{m^{\prime}}\}.\label{eq:Master}
\end{align}
To find its solution,
we use the relations between $\bar{R}_{mm^{\prime}}^{(-)}$ and $\bar{R}_{m^{\prime}m}^{(+)}$
and between $\bar{R}_{mm^{\prime}}^{(+)}$ and $\bar{R}_{m^{\prime}m}^{(-)}$.
Since $p_{p}$ is given by Eq. (\ref{eq:p_p}),
the transition rates satisfy
\begin{align}
  \hspace{-10pt}
  \frac{\bar{R}_{mm^{\prime}}^{(-)}}{\bar{R}_{m^{\prime}m}^{(+)}}
  =e^{\beta(E_{m}-E_{m^{\prime}}-\omega_{\textrm{p}})},
  \frac{\bar{R}_{mm^{\prime}}^{(+)}}{\bar{R}_{m^{\prime}m}^{(-)}}
  =e^{\beta(E_{m}-E_{m^{\prime}}+\omega_{\textrm{p}})}.\label{eq:detailed-balance}
\end{align}
[In deriving them we have used the identity
  $e^{-\beta E_{p}}\delta(\Delta E_{\mp})=e^{\beta(E_{m}-E_{m^{\prime}}\mp\omega_{\textrm{p}})}e^{-\beta E_{p^{\prime}}}\delta(\Delta E_{\mp})$.]
Equation (\ref{eq:detailed-balance})
represents the detailed balance between
magnons-pair creation and annihilation
because 
$H_{\textrm{sb}}^{(\textrm{abs})}$ and $H_{\textrm{sb}}^{(\textrm{emi})}$
describe the pair creation and annihilation, respectively. 
Combining Eq. (\ref{eq:detailed-balance}) with Eq. (\ref{eq:Master}),
we have
\begin{align}
  0=&\sum\limits_{m^{\prime}}\bar{R}_{mm^{\prime}}^{(-)}
  [\bar{p}_{m}-e^{\beta(E_{m^{\prime}}-E_{m}+\omega_{\textrm{p}})}\bar{p}_{m^{\prime}}]\notag\\
  &+\sum\limits_{m^{\prime}}\bar{R}_{mm^{\prime}}^{(+)}
  [\bar{p}_{m}-e^{\beta(E_{m^{\prime}}-E_{m}-\omega_{\textrm{p}})}\bar{p}_{m^{\prime}}].\label{eq:Master-final}
\end{align}
By assuming the $\bar{p}_{m}$ of the form
\begin{align}
  \bar{p}_{m}=\frac{e^{-\beta (E_{m}-\mu N_{m})}}
      {\sum\limits_{m^{\prime\prime}}e^{-\beta (E_{m^{\prime\prime}}-\mu N_{m^{\prime\prime}})}},\label{eq:p_m}
\end{align}
and substituting Eq. (\ref{eq:p_m}) into Eq. (\ref{eq:Master-final}),
we can show that
both terms on the right-hand side of Eq. (\ref{eq:Master-final}) are zero if
\begin{align}
  \mu=\frac{\omega_{p}}{2}.\label{eq:mu}
\end{align}
[For the first and second terms in Eq. (\ref{eq:Master-final}), 
$N_{m^{\prime}}-N_{m}=-2$ and $2$, respectively, 
because two magnons are annihilated by $H_{\textrm{sb}}^{(\textrm{emi})}$ 
and created by $H_{\textrm{sb}}^{(\textrm{abs})}$.]
We have chosen 
the chemical potentials of $\alpha$-band magnons and $\beta$-band magnons 
to be the same 
because
the change in the number of $\alpha$-band magnons due to $H_{\textrm{sb}}(t)$
is the same as the change in the number of $\beta$-band magnons.
Since the magnon operators satisfy the commutation relations for bosons,
the solution to Eq. (\ref{eq:p_m})
gives the Bose distribution function~\cite{FW}.
Indeed, we can express 
$\langle N_{\textrm{s}}\rangle=\sum_{m}N_{m}\bar{p}_{m}$ 
as the sum of the Bose distribution functions with $\mu=\omega_{p}/2$ (see Appendix G).
Thus
the magnon distribution function of our nonequilibrium steady state 
is given by the Bose distribution function with $\mu=\omega_{p}/2$.
This finite $\mu$ results from
the detailed balance of Eq. (\ref{eq:detailed-balance}).

To obtain a deeper understanding of our mechanism for generating the $\mu$,
we remark on some of the properties of Eqs. (\ref{eq:R-bar^-}) and (\ref{eq:R-bar^+}).
The $\bar{R}_{mm^{\prime}}^{(-)}$ in Eq. (\ref{eq:R-bar^-}) includes
the factor
$|\langle m^{\prime}|\langle p^{\prime}|H_{\textrm{sb}}^{(\textrm{emi})}|p\rangle|m\rangle|^{2}\delta(\Delta E_{-})$;
the $\bar{R}_{mm^{\prime}}^{(+)}$ in Eq. (\ref{eq:R-bar^+}) includes
the factor
$|\langle m^{\prime}|\langle p^{\prime}|H_{\textrm{sb}}^{(\textrm{abs})}|p\rangle|m\rangle|^{2}\delta(\Delta E_{+})$.
The former factor is finite only if
\begin{align}
  \Delta E_{-}=E_{m}+E_{p}-E_{m^{\prime}}-E_{p^{\prime}}-\omega_{\textrm{p}}=0;\label{eq:cond_R^-}
\end{align}
the latter is finite only if
\begin{align}
  \Delta E_{+}=E_{m}+E_{p}-E_{m^{\prime}}-E_{p^{\prime}}+\omega_{\textrm{p}}=0.\label{eq:cond_R^+}
\end{align}
A detailed examination of these conditions is helpful in obtaining 
the deeper understanding of our mechanism.
Since $H_{\textrm{sb}}^{(\textrm{emi})}$ is given by Eq. (\ref{eq:Hsb^emi}),
we can express Eq. (\ref{eq:cond_R^-}) as
\begin{align}
  E_{m}(N_{m})+E_{p}-E_{m^{\prime}}(N_{m}-2)-E_{p^{\prime}}-\omega_{\textrm{p}}=0,\label{eq:cond_R^-_rewr}
\end{align}
where we have explicitly written the magnon numbers
for the states $|m\rangle$ and $|m^{\prime}\rangle$.
For the scattering processes due to the $H_{\textrm{sb}}^{(\textrm{emi})}$
we have
\begin{align}
  E_{m}(N_{m})-E_{m^{\prime}}(N_{m}-2)
  &\approx E_{m}(N_{m})-E_{m^{\prime}}(N_{m})+2\mu \notag\\
  &=\epsilon_{\alpha}(\boldq)+\epsilon_{\beta}(\boldq^{\prime})+2\mu,\label{eq:cond_R^-_rewr_1}
\end{align}
and
\begin{align}
  E_{p^{\prime}}-E_{p}\approx \omega_{\textrm{p}}.\label{eq:cond_R^-_rewr_2}
\end{align}
Thus Eq. (\ref{eq:cond_R^-}) is divided into 
$\epsilon_{\alpha}(\boldq)+\epsilon_{\beta}(\boldq^{\prime})=\omega_{\textrm{p}}$
and $2\mu=\omega_{\textrm{p}}$.
Similarly, we can divide Eq. (\ref{eq:cond_R^+}) into the same two equations.
Therefore 
both the change in the magnon number and 
the term $(E_{p}-E_{p^{\prime}})$ are necessary 
for obtaining the finite $\mu$.
The term $(E_{p}-E_{p^{\prime}})$ appears
only if the time-periodic magnetic field is treated as the quantized field.
[If it is treated as the classical field,
  that term is absent because of lack of the creation or annihilation operator of a photon;
  in this classical case,
  the corresponding conditions 
  might be $E_{m}(N_{m})-E_{m^{\prime}}(N_{m}-2)-\omega_{\textrm{p}}=0$
  and $E_{m}(N_{m})-E_{m^{\prime}}(N_{m}+2)+\omega_{\textrm{p}}=0$, 
  and thus the $\mu$ should be zero.] 
We thus conclude that
the quantum-mechanical treatment of the time-periodic magnetic field 
and the Markovian-bath treatment of its effects on the magnon system are 
essential for obtaining the finite $\mu$ in the nonequilibrium steady state. 

\section{Discussion}

We first compare our results with experimental results.
Experimental studies of the parametric parallel pumping of YIG~\cite{MagBEC-Nature,Estimate-MagChem}
have shown that
after a certain period of time under the time-periodic magnetic field, 
the magnon distribution function can be approximated by
the Bose distribution function with finite $\mu$. 
This means that 
the time-periodic magnetic field generates $\mu$ 
because the zero of this $\mu$ is set to the value without it.
Our result can qualitatively explain this experimental result.
However, there is a quantitative difference between them
because 
the experimentally estimated value of $\mu$
reaches $\mu\approx \omega_{\textrm{p}}/4$ for some pumping powers~\cite{Estimate-MagChem}.
Although a quantitatively appropriate theoretical description
is beyond the scope of the present study,
we believe that
for the quantitative comparison with the experimental results
the effect of a phonon should be taken into account.
This is because
the phonon-assisted processes,
which are similar to the indirect transitions~\cite{Ziman,Ashcroft}
in semiconductors,
may be vital for understanding
how a pair of magnons in different bands is created or annihilated
by a GHz-frequency photon.
It is known that
in order to describe the optical properties of semiconductors,
one needs to consider not only the direct transitions,
the transitions using only a photon, 
but also the indirect transitions,
the transitions using a photon and a phonon~\cite{Ziman,Ashcroft}.
Such phonon-assisted processes can be used even for the optical properties of magnon systems.
If the energy of a phonon is set to $0.03$ eV~\cite{Ziman}, 
the sum of it and the energy of a GHz-frequency photon 
is comparable with
the energy of a pair of small-$|\boldq|$ magnons in the lowest branch and
the second lowest branch for YIG.
Note that the energy of a small-$|\boldq|$ magnon in the second lowest branch is
about $7$THz$\approx 0.03$eV~\cite{Bauer-Ferri},
where we have used $1$THz$\approx 4$meV,
the relation between frequency units and energy units
used in the neutron scattering experiments~\cite{Neutron-YIG} for YIG. 

We then discuss the differences between the standard theory and our theory.
As described in Sect. 1,
the time-periodic magnetic field is treated as a classical field
in the standard theory~\cite{S-theory1,S-theory2,S-theory3,PRB-Stheory1,Inv-Stheory}.
Because of this treatment,
the standard theory uses 
an approximation whose validity is uncertain:
the factor $\cos(\omega_{\textrm{p}}t)$ of
$H_{\textrm{pump}}(t)
=g\mu_{\textrm{B}}h\cos(\omega_{\textrm{p}}t)\sum\textstyle_{\boldj}S^{z}_{\boldj}$
is replaced by $e^{-i\omega_{\textrm{p}}t}$ or $e^{i\omega_{\textrm{p}}t}$ 
for the magnons-pair creation or annihilation term,
respectively~\cite{PRB-Stheory1,Inv-Stheory}. 
In contrast,
our theory does not use that approximation
because such exponential time dependence appears naturally
in the quantized magnetic field.
This difference is one advantage of our theory.
Another advantage is the presence of a photon bath.
Since the standard theory~\cite{S-theory1,S-theory2,S-theory3,PRB-Stheory1,Inv-Stheory}
does not consider a photon bath,
magnon-number conservation is always violated by 
the magnons-pair creation and annihilation terms due to the time-periodic magnetic field, 
and, as a result, $\mu=0$~\cite{S-theory2,Inv-Stheory}.
In our theory
the rates of the pair creation and the pair annihilation satisfy the detailed balance
in the nonequilibrium steady state,
and, as a result,
the effects of their terms are reduced to $\mu=\omega_{\textrm{p}}/2$.

We now discuss
the implications of our theory.
The framework of our master equation
is applicable to other collinear magnets,
in which the magnetization directions are collinear, 
because in a similar way 
$H_{\textrm{sb}}(t)$ can be expressed
as the magnons-pair creation and annihilation terms. 
Thus, even for other collinear magnets,
the distribution function of nonequilibrium steady-state magnons
in the parametric parallel pumping
could be approximated by the Bose distribution function with finite $\mu$. 
Since our theory can be extended
to a more complicated model of YIG~\cite{YIG-review,YIG-1stPrinciple},
our theory provides an important step towards a thorough understanding of
properties of nonequilibrium magnons of YIG. 
In addition,
since the similar mechanism
can be used to generate $\mu$ for antiferromagnets, 
our results will stimulate further research of
the parametric parallel pumping and the magnon BEC for antiferromagnets.
It should be noted that
for the parametric parallel pumping of an antiferromagnet 
a pair of magnons in different bands can be created or annihilated by a GHz-frequency photon
even without the assistance of a phonon
because the band splitting is induced by 
the Zeeman energy of the time-independent magnetic field~\cite{NA-AF}
and it is much smaller than that induced by the Heisenberg exchange interaction.
This property is distinct from the property for ferrimagnets,
and thus may be an advantage of antiferromagnets. 

\section{Summary}

We have studied
the magnon properties of the two-sublattice ferrimagnet
in the nonequilibrium steady state under the time-periodic magnetic field.
We have introduced the model Hamiltonian, in which 
the magnon system in the parametric parallel pumping
is described by the system of magnons with the weak coupling to the Markovian bath of photons. 
To understand
the nonequilibrium steady-state properties of this system,
we have derived the master equation of the reduced density matrix of the magnons,
and then we have studied its steady-state solution.
We have shown that
the magnon distribution function of the nonequilibrium steady state
becomes the Bose distribution function with $\mu=\omega_{\textrm{p}}/2$.
This result can qualitatively explain the generation of the magnon chemical potential
in experiments~\cite{MagBEC-Nature,Estimate-MagChem}, 
and it is distinct from
the value of the standard theory, $\mu=0$.

\begin{acknowledgments}
    The author thanks E. Saitoh and H. Adachi for useful discussions
    about magnon properties in the parametric parallel pumping.
    This work was supported by JSPS KAKENHI Grant Number JP19K14664.
\end{acknowledgments}

\appendix

\section{Effect of the $h_{0}$ on the lowest energy of the magnon bands}

In this Appendix
we discuss the effect of the $h_{0}$
on the lowest energy of the magnon bands. 
As a concrete example
we consider the case of $S_{A} < S_{B}$ for our two-sublattice ferrimagnet. 
In this case 
we take $h_{0} > 0$ because 
the $M$ satisfies $M > 0$ [see Eq. (\ref{eq:Magnetization})].
As a result, 
the $-h_{0}M$ term in Eq. (\ref{eq:Hs-sub})
makes the directions of the time-independent magnetic field and the magnetization parallel. 
Then,
from Eqs. (\ref{eq:E_alpha}){--}(\ref{eq:DelE}),
we see that 
the lowest energy in $\tilde{\epsilon}_{\alpha}(\boldq)$ is given by
\begin{align}
  \tilde{\epsilon}_{\alpha}(\boldzero)=2(S_{B}-S_{A})J(\boldzero)-g\mu_{\textrm{B}}h_{0},
\end{align}
and that in $\tilde{\epsilon}_{\beta}(\boldq)$ is given by
\begin{align}
  \tilde{\epsilon}_{\beta}(\boldzero)=g\mu_{\textrm{B}}h_{0}.
\end{align}
Since $2(S_{B}-S_{A})J(\boldzero)$ is usually larger than $g\mu_{\textrm{B}}h_{0}$,
the lowest energy for $S_{A} < S_{B}$
is $\tilde{\epsilon}_{\beta}(\boldzero)=g\mu_{\textrm{B}}h_{0}$.
Thus the $h_{0}$ makes
the lowest energy of the magnon bands nonzero.
The case of $S_{A} > S_{B}$ can be discussed in a similar way.

\section{Derivation of Eqs. (\ref{eq:Hsb}){--}(\ref{eq:Hsb^emi})}

In this Appendix we derive Eqs. (\ref{eq:Hsb}){--}(\ref{eq:Hsb^emi}).
By substituting Eqs. (\ref{eq:Bogo1}) and (\ref{eq:Bogo2}) 
into Eq. (\ref{eq:Hsb-ab}),
we can rewrite $H_{\textrm{sb}}(t)$ as follows:
\begin{align}
  H_{\textrm{sb}}(t)=
  &e^{-i\omega_{\textrm{p}}t}\sum\limits_{\boldq,\boldq^{\prime},\lambda}
  \tilde{C}_{\boldq-\boldq^{\prime}\lambda}c_{\boldq-\boldq^{\prime}\lambda}\notag\\
  \times
  &[B_{\boldq\boldq^{\prime}}
  (\alpha_{\boldq}^{\dagger}\beta_{\boldq^{\prime}}^{\dagger}
  -\beta_{\boldq}\alpha_{\boldq^{\prime}})
  -A_{\boldq\boldq^{\prime}}
  (\alpha_{\boldq}^{\dagger}\alpha_{\boldq^{\prime}}
  -\beta_{\boldq^{\prime}}^{\dagger}\beta_{\boldq})]\notag\\  
  +&e^{i\omega_{\textrm{p}}t}\sum\limits_{\boldq,\boldq^{\prime},\lambda}
  \tilde{C}_{\boldq-\boldq^{\prime}\lambda}^{\ast}c_{\boldq-\boldq^{\prime}\lambda}^{\dagger}\notag\\
  \times 
  &[B_{\boldq\boldq^{\prime}}
  (\beta_{\boldq^{\prime}}\alpha_{\boldq}
  -\alpha_{\boldq^{\prime}}^{\dagger}\beta_{\boldq}^{\dagger})
  -A_{\boldq\boldq^{\prime}}
  (\alpha_{\boldq^{\prime}}^{\dagger}\alpha_{\boldq}
  -\beta_{\boldq}^{\dagger}\beta_{\boldq^{\prime}})],\label{eq:Hsb-first}
\end{align}
where $B_{\boldq\boldq^{\prime}}$ is given by Eq. (\ref{eq:Bqq'}),
and $A_{\boldq\boldq^{\prime}}$ is given by 
\begin{align}
  A_{\boldq\boldq^{\prime}}=
 \cosh\theta_{\boldq}\cosh\theta_{\boldq^{\prime}}
  -\sinh\theta_{\boldq}\sinh\theta_{\boldq^{\prime}}.
\end{align}
Because of energy and momentum conservation
the relevant terms of Eq. (\ref{eq:Hsb-first}) are given by
Eqs. (\ref{eq:Hsb}){--}(\ref{eq:Hsb^emi})
because 
the single-magnon excitation terms in Eq. (\ref{eq:Hsb-first}),
the terms including $A_{\boldq\boldq^{\prime}}$, 
are irrelevant~\cite{Hmp}.

\section{Derivation of Eq. (\ref{eq:approx})}

In this Appendix
we explain the details of the derivation of Eq. (\ref{eq:approx}).
We first derive a general expression of the equation of motion of $\rho_{\textrm{s}}(t)$,
and then rewrite it by using the Born-Markov approximation,
which is valid for a system with weak coupling to a Markovian bath.
The following derivation is based on
the derivation described in Ref. \cite{Kubo-text}.  

First, we rewrite Eq. (\ref{eq:Liouville})
as the equation of motion of $\rho_{\textrm{s}}(t)$.
To do this,
we introduce projection operators $\mathcal{P}$ and $\mathcal{P}^{\prime}$,
\begin{align}
  &\mathcal{P}=\rho_{\textrm{b}}\textrm{tr}_{\textrm{b}},\label{eq:P}\\
  &\mathcal{P}^{\prime}=1-\mathcal{P},\label{eq:P'}
\end{align}
where $\rho_{\textrm{b}}$ is the density matrix of photons,
\begin{align}
  \rho_{\textrm{b}}=\frac{e^{-\beta H_{\textrm{b}}}}{\textrm{tr}_{\textrm{b}}e^{-\beta H_{\textrm{b}}}},
  \label{eq:rho_b}
\end{align}
and $\beta=(k_{\textrm{B}}T)^{-1}$.
Since $\rho(t)=\mathcal{P}\rho(t)+\mathcal{P}^{\prime}\rho(t)$, 
we can rewrite Eq. (\ref{eq:Liouville}) as
a set of the equations of motion of $\mathcal{P}\rho(t)$ and
$\mathcal{P}^{\prime}\rho(t)$; the results are
\begin{align}
  &\frac{d}{dt}\mathcal{P}\rho(t)
  =\mathcal{P}\mathcal{L}\mathcal{P}\rho(t)
  +\mathcal{P}\mathcal{L}\mathcal{P}^{\prime}\rho(t),\label{eq:Prho}\\
  &\frac{d}{dt}\mathcal{P}^{\prime}\rho(t)
  =\mathcal{P}^{\prime}\mathcal{L}\mathcal{P}\rho(t)
  +\mathcal{P}^{\prime}\mathcal{L}\mathcal{P}^{\prime}\rho(t),\label{eq:P'rho}
\end{align}
where $\mathcal{L}$ is the Liouville operator for $H$, 
\begin{align}
  \mathcal{L}\rho(t)=\frac{1}{i}[H,\rho(t)].
\end{align}
In deriving Eqs. (\ref{eq:Prho}) and (\ref{eq:P'rho})
we have used the identities $\mathcal{P}^{2}=\mathcal{P}$
and $\mathcal{P}^{\prime}\mathcal{P}=\mathcal{P}\mathcal{P}^{\prime}=0$.
Then 
the formal solution to Eq. (\ref{eq:P'rho}) is given by
\begin{align}
  \mathcal{P}^{\prime}\rho(t)
  =\int_{0}^{t}d\tau
  e^{(t-\tau)\mathcal{P}^{\prime}\mathcal{L}}\mathcal{P}^{\prime}
  \mathcal{L}\mathcal{P}\rho(\tau).\label{eq:P'rho-FormSol}
\end{align}
Here we have supposed that 
$\rho(0)=\rho_{\textrm{b}}\rho_{\textrm{s}}(0)$;
because of this initial-state condition, $\mathcal{P}^{\prime}\rho(0)=0$.
Substituting Eq. (\ref{eq:P'rho-FormSol}) into the second term
on the right-hand side of Eq. (\ref{eq:Prho}), we have
\begin{align}
  \frac{d}{dt}\mathcal{P}\rho(t)
  =\mathcal{P}\mathcal{L}\mathcal{P}\rho(t)
  +\mathcal{P}\mathcal{L}\int_{0}^{t}d\tau
  e^{(t-\tau)\mathcal{P}^{\prime}\mathcal{L}}\mathcal{P}^{\prime}
  \mathcal{L}\mathcal{P}\rho(\tau).\label{eq:Prho-next}
\end{align}
This equation can be rewritten as the equation of motion of $\rho_{\textrm{s}}(t)$
because
\begin{align}
  \mathcal{P}\rho(t)
  =\rho_{\textrm{b}}\textrm{tr}_{\textrm{b}}\rho(t)
  =\rho_{\textrm{b}}\rho_{\textrm{s}}(t).
\end{align}
As we derive in Appendix D, 
we obtain
\begin{align}
  \frac{d\rho_{\textrm{s}}(t)}{dt}
  =\mathcal{L}_{\textrm{s}}\rho_{\textrm{s}}(t)
  +\textrm{tr}_{\textrm{b}}
  \mathcal{L}_{\textrm{sb}}
  \int_{0}^{t}d\tau
  e^{(t-\tau)\mathcal{P}^{\prime}\mathcal{L}\mathcal{P}^{\prime}}
  \mathcal{L}_{\textrm{sb}}
  \rho_{\textrm{b}}\rho_{\textrm{s}}(\tau).\label{eq:rho_m}
\end{align}
In deriving this equation
we have introduced the Liouville operators for $H_{\textrm{s}}$,   
$H_{\textrm{sb}}(t)$, and $H_{\textrm{b}}$ as follows:
\begin{align}
  &\mathcal{L}_{\textrm{s}}\rho(t)
  =\frac{1}{i}[H_{\textrm{s}},\rho(t)],\label{eq:Liouville_s}\\
  &\mathcal{L}_{\textrm{sb}}\rho(t)
  =\frac{1}{i}[H_{\textrm{sb}}(t),\rho(t)],\label{eq:Liouville_sb}\\
  &\mathcal{L}_{\textrm{b}}\rho(t)
  =\frac{1}{i}[H_{\textrm{b}},\rho(t)],\label{eq:Liouville_b}
\end{align}
where
\begin{align}
  \mathcal{L}=
  \mathcal{L}_{\textrm{s}}+\mathcal{L}_{\textrm{sb}}
  +\mathcal{L}_{\textrm{b}}.\label{eq:Liouville_tot}
\end{align}

Then we can write Eq. (\ref{eq:rho_m}) in a simpler form
by using the Born-Markov approximation.
This approximation 
is appropriate for a system with weak coupling to a Markovian bath,
and it consists of two approximations. 
The first approximation is similar to 
the Born approximation for the scattering theory of electrons.
Since the second term on the right-hand side of Eq. (\ref{eq:rho_m})
has two $\mathcal{L}_{\textrm{sb}}$'s,
corresponding to two $H_{\textrm{sb}}(t)$'s [Eq. (\ref{eq:Liouville_sb})], 
we can replace
$\mathcal{L}=\mathcal{L}_{\textrm{s}}+\mathcal{L}_{\textrm{sb}}+\mathcal{L}_{\textrm{b}}$
of $e^{(t-\tau)\mathcal{P}^{\prime}\mathcal{L}\mathcal{P}^{\prime}}$ in that term  
by $\mathcal{L}_{0}=\mathcal{L}_{\textrm{s}}+\mathcal{L}_{\textrm{b}}$
by using the second-order perturbation theory for $H_{\textrm{sb}}(t)$.
In addition,
since 
$\mathcal{P}^{\prime}\mathcal{L}_{0}\mathcal{P}^{\prime}=\mathcal{P}^{\prime}\mathcal{L}_{0}-\mathcal{P}^{\prime}\mathcal{L}_{0}\mathcal{P}=\mathcal{P}^{\prime}\mathcal{L}_{0}$, 
we have $e^{(t-\tau)\mathcal{P}^{\prime}\mathcal{L}_{0}\mathcal{P}^{\prime}}
=e^{(t-\tau)\mathcal{P}^{\prime}\mathcal{L}_{0}}=\mathcal{P}^{\prime}e^{(t-\tau)\mathcal{L}_{0}}$. 
Combining those results with Eq. (\ref{eq:rho_m}),
we obtain
\begin{align}
  \frac{d\rho_{\textrm{s}}(t)}{dt}
  =\mathcal{L}_{\textrm{s}}\rho_{\textrm{s}}(t)
  +\textrm{tr}_{\textrm{b}}
  \mathcal{L}_{\textrm{sb}}
  \int_{0}^{t}d\tau
  e^{(t-\tau)\mathcal{L}_{0}}
  \mathcal{L}_{\textrm{sb}}
  \rho_{\textrm{b}}\rho_{\textrm{s}}(\tau),\label{eq:rho_m-Born}
\end{align}
where we have used
$\mathcal{P}\mathcal{L}_{\textrm{sb}}\mathcal{P}^{\prime}=\mathcal{P}\mathcal{L}_{\textrm{sb}}$,
which results in   
$\textrm{tr}_{\textrm{b}}\mathcal{L}_{\textrm{sb}}\mathcal{P}^{\prime}
=\rho_{\textrm{b}}^{-1}\mathcal{P}\mathcal{L}_{\textrm{sb}}\mathcal{P}^{\prime}
=\rho_{\textrm{b}}^{-1}\mathcal{P}\mathcal{L}_{\textrm{sb}}
=\textrm{tr}_{\textrm{b}}\mathcal{L}_{\textrm{sb}}$.
The second approximation is the Markov approximation,
which is valid for a Markovian bath.
To use it,
we rewrite Eq. (\ref{eq:rho_m-Born})
in the interaction picture. 
First,
by using Eqs. (\ref{eq:Liouville_s}){--}(\ref{eq:Liouville_b}),
we can express Eq. (\ref{eq:rho_m-Born}) as follows:
\begin{align}
  \frac{d\rho_{\textrm{s}}(t)}{dt}
  =\frac{1}{i}[H_{\textrm{s}},\rho_{\textrm{s}}(t)]
  -\textrm{tr}_{\textrm{b}}
  [H_{\textrm{sb}}(t),F(t)].\label{eq:rho_m-Born-rewrite1}
\end{align}
where
\begin{align}
  F(t)=\int_{0}^{t}d\tau
    e^{-i(t-\tau)H_{0}}
  [H_{\textrm{sb}}(\tau),\rho_{\textrm{b}}\rho_{\textrm{s}}(\tau)]
  e^{i(t-\tau)H_{0}},
\end{align}
and
\begin{align}
  H_{0}=H_{\textrm{s}}+H_{\textrm{b}}.
\end{align}
[Note that because of Eqs. (\ref{eq:Liouville_s}) and (\ref{eq:Liouville_b})
$\mathcal{L}_{0}=\mathcal{L}_{\textrm{s}}+\mathcal{L}_{\textrm{b}}$ satisfies
$e^{\mathcal{L}_{0}t}(\cdots)=e^{-iH_{0}t}(\cdots)e^{iH_{0}t}$.]
Then,
by using the operators in the interaction picture,
i.e., Eqs. (\ref{eq:rho_s^I}) and  (\ref{eq:H_sb^I}), 
we can rewrite Eq. (\ref{eq:rho_m-Born-rewrite1}) in the form
\begin{align}
  \frac{d\rho_{\textrm{s}}^{(\textrm{I})}(t)}{dt}
  =-\textrm{tr}_{\textrm{b}}
  [H_{\textrm{sb}}^{(\textrm{I})}(t),
  \int_{0}^{t}d\tau
      [H_{\textrm{sb}}^{(\textrm{I})}(\tau),
        \rho_{\textrm{b}}\rho_{\textrm{s}}^{(\textrm{I})}(\tau)]].\label{eq:rho_m-Born-rewrite2}
\end{align}
We suppose that
the time variation of $\rho_{\textrm{s}}^{(\textrm{I})}(t)$,
which is characterized by $\tau_{\textrm{r}}$,
is slower than that of $H_{\textrm{sb}}^{(\textrm{I})}(t)$.
(This condition is satisfied
for a system with weak coupling to a Markovian bath.)
Because of this,
we can approximate
$\rho_{\textrm{s}}^{(\textrm{I})}(\tau)$ in Eq. (\ref{eq:rho_m-Born-rewrite2}) as
$\rho_{\textrm{s}}^{(\textrm{I})}(t)$;
thus, Eq. (\ref{eq:rho_m-Born-rewrite2}) becomes 
\begin{align}
  \frac{d\rho_{\textrm{s}}^{(\textrm{I})}(t)}{dt}
  =-\textrm{tr}_{\textrm{b}}
  [H_{\textrm{sb}}^{(\textrm{I})}(t),
  \int_{0}^{t}d\tau
      [H_{\textrm{sb}}^{(\textrm{I})}(\tau),
        \rho_{\textrm{b}}\rho_{\textrm{s}}^{(\textrm{I})}(t)]].\label{eq:rho_m-BornMarkov}
\end{align}

\section{Derivation of Eq. (\ref{eq:rho_m})}

In this Appendix we derive Eq. (\ref{eq:rho_m}).
This derivation consists of the following three steps. 

First,
we rewrite the first term on the right-hand side of Eq. (\ref{eq:Prho-next}). 
By using Eq. (\ref{eq:Liouville_tot}),
the first term can be expressed as
\begin{align}
  \mathcal{P}\mathcal{L}\mathcal{P}\rho(t)
  =\mathcal{P}\mathcal{L}_{0}\mathcal{P}\rho(t)
  +\mathcal{P}\mathcal{L}_{\textrm{sb}}\mathcal{P}\rho(t),\label{eq:AppC-1}
\end{align}
where $\mathcal{L}_{0}=\mathcal{L}_{\textrm{s}}+\mathcal{L}_{\textrm{b}}$.
Since $\mathcal{P}\rho(t)=\rho_{\textrm{b}}\rho_{\textrm{s}}(t)$
follows from Eqs. (\ref{eq:rho_s}) and (\ref{eq:P}),
the first term of Eq. (\ref{eq:AppC-1}) becomes
\begin{align}
  \mathcal{P}\mathcal{L}_{0}\rho_{\textrm{b}}\rho_{\textrm{s}}(t)
  =&\mathcal{P}\mathcal{L}_{\textrm{s}}\rho_{\textrm{b}}\rho_{\textrm{s}}(t)\notag\\
  =&\mathcal{L}_{\textrm{s}}\rho_{\textrm{b}}\rho_{\textrm{s}}(t),\label{eq:AppC-2}
\end{align}
where we have used $\mathcal{L}_{\textrm{b}}\rho_{\textrm{b}}=0$.
In addition,
the second term of Eq. (\ref{eq:AppC-1}) becomes
\begin{align}
  \mathcal{P}\mathcal{L}_{\textrm{sb}}\rho_{\textrm{b}}\rho_{\textrm{s}}(t)
  =&\rho_{\textrm{b}}\textrm{tr}_{\textrm{b}}
  \frac{1}{i}[H_{\textrm{sb}}(t),\rho_{\textrm{b}}\rho_{\textrm{s}}(t)]\notag\\
  =&\frac{1}{i}\rho_{\textrm{b}}
  [\textrm{tr}_{\textrm{b}}H_{\textrm{sb}}(t)\rho_{\textrm{b}},\rho_{\textrm{s}}(t)]\notag\\
  =&0,\label{eq:AppC-3}
\end{align}
where we have used $\textrm{tr}_{\textrm{b}}H_{\textrm{sb}}(t)\rho_{\textrm{b}}=0$.
Combining Eqs. (\ref{eq:AppC-2}) and (\ref{eq:AppC-3})
with Eq. (\ref{eq:AppC-1}),
we have
\begin{align}
  \mathcal{P}\mathcal{L}\mathcal{P}\rho(t)
  =\mathcal{L}_{\textrm{s}}\rho_{\textrm{b}}\rho_{\textrm{s}}(t).\label{eq:AppC-4}
\end{align}

Next, 
we rewrite the second term of Eq. (\ref{eq:Prho-next}) in a similar manner.
The $\mathcal{P}^{\prime}\mathcal{L}\mathcal{P}$, which appears in that term,
can be expressed as follows: 
\begin{align}
  \mathcal{P}^{\prime}\mathcal{L}\mathcal{P}
  =&\mathcal{P}^{\prime}\mathcal{L}_{0}\mathcal{P}
  +\mathcal{P}^{\prime}\mathcal{L}_{\textrm{sb}}\mathcal{P}\notag\\
  =&\mathcal{P}^{\prime}\mathcal{L}_{\textrm{sb}}\mathcal{P}\notag\\
  =&\mathcal{L}_{\textrm{sb}}\mathcal{P}.\label{eq:AppC-5}
\end{align}
Here we have used
$\mathcal{P}^{\prime}\mathcal{L}_{0}\mathcal{P}=0$ 
and $\mathcal{P}\mathcal{L}_{\textrm{sb}}\mathcal{P}=0$,
which follow
from $\mathcal{P}\mathcal{L}_{0}\mathcal{P}=\mathcal{L}_{0}\mathcal{P}$ and 
from $\textrm{tr}_{\textrm{b}}H_{\textrm{sb}}(t)\rho_{\textrm{b}}=0$,
respectively.
Thus 
the second term of Eq. (\ref{eq:Prho-next}) becomes
\begin{align}
  \mathcal{P}\mathcal{L}\int_{0}^{t}d\tau
  e^{(t-\tau)\mathcal{P}^{\prime}\mathcal{L}}\mathcal{P}^{\prime}
  \mathcal{L}\mathcal{P}\rho(\tau)
  =&
  \mathcal{P}\mathcal{L}\int_{0}^{t}d\tau
  e^{(t-\tau)\mathcal{P}^{\prime}\mathcal{L}}\notag\\
  &\times 
  \mathcal{L}_{\textrm{sb}}\rho_{\textrm{b}}\rho_{\textrm{s}}(\tau).\label{eq:AppC-6-prepre}
\end{align}
Using $\mathcal{P}^{\prime}=1-\mathcal{P}$, $\mathcal{P}^{\prime}\mathcal{P}=0$,
and Eq. (\ref{eq:AppC-5}), 
we can express part of Eq. (\ref{eq:AppC-6-prepre}) as follows: 
\begin{align}
  e^{(t-\tau)\mathcal{P}^{\prime}\mathcal{L}}
  \mathcal{L}_{\textrm{sb}}\rho_{\textrm{b}}\rho_{\textrm{s}}(\tau)
  =&e^{(t-\tau)\mathcal{P}^{\prime}\mathcal{L}\mathcal{P}^{\prime}}e^{(t-\tau)\mathcal{P}^{\prime}\mathcal{L}\mathcal{P}}
  \mathcal{L}_{\textrm{sb}}\rho_{\textrm{b}}\rho_{\textrm{s}}(\tau)\notag\\  
  =&e^{(t-\tau)\mathcal{P}^{\prime}\mathcal{L}\mathcal{P}^{\prime}}
  e^{(t-\tau)\mathcal{P}^{\prime}\mathcal{L}_{\textrm{sb}}\mathcal{P}}
  \mathcal{L}_{\textrm{sb}}\rho_{\textrm{b}}\rho_{\textrm{s}}(\tau)\notag\\
  =&e^{(t-\tau)\mathcal{P}^{\prime}\mathcal{L}\mathcal{P}^{\prime}}
  \mathcal{L}_{\textrm{sb}}\rho_{\textrm{b}}\rho_{\textrm{s}}(\tau).\label{eq:AppC-6-pre}
\end{align}
In deriving the final line 
we have used
\begin{align}
  e^{(t-\tau)\mathcal{P}^{\prime}\mathcal{L}_{\textrm{sb}}\mathcal{P}}
  \mathcal{L}_{\textrm{sb}}\rho_{\textrm{b}}\rho_{\textrm{s}}(\tau)
  =&[1+(t-\tau)\mathcal{P}^{\prime}\mathcal{L}_{\textrm{sb}}\mathcal{P}]
  \mathcal{L}_{\textrm{sb}}\mathcal{P}\rho(\tau)\notag\\
  =&\mathcal{L}_{\textrm{sb}}\rho_{\textrm{b}}\rho_{\textrm{s}}(\tau),
\end{align}
where $\mathcal{P}^{\prime}\mathcal{P}=0$ and $\mathcal{P}\mathcal{L}_{\textrm{sb}}\mathcal{P}=0$.
Combining Eq. (\ref{eq:AppC-6-pre}) with Eq. (\ref{eq:AppC-6-prepre}),
we obtain
\begin{align}
  &\mathcal{P}\mathcal{L}\int_{0}^{t}d\tau
  e^{(t-\tau)\mathcal{P}^{\prime}\mathcal{L}}
  \mathcal{L}_{\textrm{sb}}\rho_{\textrm{b}}\rho_{\textrm{s}}(\tau)\notag\\
  =&\mathcal{P}\mathcal{L}\int_{0}^{t}d\tau
  e^{(t-\tau)\mathcal{P}^{\prime}\mathcal{L}\mathcal{P}^{\prime}}
  \mathcal{L}_{\textrm{sb}}\rho_{\textrm{b}}\rho_{\textrm{s}}(\tau)\notag\\
  =&\mathcal{P}\mathcal{L}\mathcal{P}^{\prime}\int_{0}^{t}d\tau
  e^{(t-\tau)\mathcal{P}^{\prime}\mathcal{L}\mathcal{P}^{\prime}}
  \mathcal{L}_{\textrm{sb}}\rho_{\textrm{b}}\rho_{\textrm{s}}(\tau)\notag\\
  &+\mathcal{P}\mathcal{L}\mathcal{P}\int_{0}^{t}d\tau
  e^{(t-\tau)\mathcal{P}^{\prime}\mathcal{L}\mathcal{P}^{\prime}}
  \mathcal{L}_{\textrm{sb}}\rho_{\textrm{b}}\rho_{\textrm{s}}(\tau)\notag\\
  =&\mathcal{P}\mathcal{L}\mathcal{P}^{\prime}\int_{0}^{t}d\tau
  e^{(t-\tau)\mathcal{P}^{\prime}\mathcal{L}\mathcal{P}^{\prime}}
  \mathcal{L}_{\textrm{sb}}\rho_{\textrm{b}}\rho_{\textrm{s}}(\tau)\notag\\
  =&\mathcal{P}\mathcal{L}_{\textrm{sb}}\int_{0}^{t}d\tau
  e^{(t-\tau)\mathcal{P}^{\prime}\mathcal{L}\mathcal{P}^{\prime}}
  \mathcal{L}_{\textrm{sb}}\rho_{\textrm{b}}\rho_{\textrm{s}}(\tau).\label{eq:AppC-6}
\end{align}
Here we have used 
$\mathcal{P}^{\prime}\mathcal{P}=0$, $\mathcal{P}\mathcal{L}_{\textrm{sb}}\mathcal{P}=0$,
and 
\begin{align}
\mathcal{P}\mathcal{L}\mathcal{P}^{\prime}
=&\mathcal{P}\mathcal{L}_{\textrm{sb}}\mathcal{P}^{\prime}\notag\\
=&\mathcal{P}\mathcal{L}_{\textrm{sb}}.\label{eq:AppC-6-next} 
\end{align}
[Equation (\ref{eq:AppC-6-next}) is derived in a similar way to Eq. (\ref{eq:AppC-5}).]

Finally,
we combine these results with Eq. (\ref{eq:Prho-next}). 
Combining Eqs. (\ref{eq:AppC-4}) and (\ref{eq:AppC-6}) with Eq. (\ref{eq:Prho-next}),
we obtain
\begin{align}
  &\frac{d}{dt}\rho_{\textrm{b}}\rho_{\textrm{s}}(t)
  =\mathcal{L}_{\textrm{s}}\rho_{\textrm{b}}\rho_{\textrm{s}}(t)\notag\\
  &+\rho_{\textrm{b}}\textrm{tr}_{\textrm{b}}\mathcal{L}_{\textrm{sb}}\int_{0}^{t}d\tau
  e^{(t-\tau)\mathcal{P}^{\prime}\mathcal{L}\mathcal{P}^{\prime}}
  \mathcal{L}_{\textrm{sb}}\rho_{\textrm{b}}\rho_{\textrm{s}}(\tau).
\end{align}
This is reduced to Eq. (\ref{eq:rho_m})
because $\mathcal{L}_{\textrm{s}}\rho_{\textrm{b}}=\rho_{\textrm{b}}\mathcal{L}_{\textrm{s}}$. 

\section{Derivation of Eqs. (\ref{eq:Master-tdep-pre}) and (\ref{eq:R-pre})}

In this Appendix we derive Eqs. (\ref{eq:Master-tdep-pre}) and (\ref{eq:R-pre}).
Since the eigenvector of $H_{\textrm{s}}$, $|m\rangle$, satisfies
\begin{align}
  H_{\textrm{s}}|m\rangle =i\frac{d}{dt}|m\rangle,
\end{align}
$p_{m}(t)=\langle m|\rho_{\textrm{s}}(t)|m\rangle$ satisfies
\begin{align}
  \frac{d p_{m}(t)}{dt}
  &=\langle m|\frac{d \rho_{\textrm{s}}(t)}{dt}|m\rangle
  -\frac{1}{i}\langle m|[H_{\textrm{s}},\rho_{\textrm{s}}(t)]|m\rangle\notag\\
  &=\langle m|\frac{d \rho_{\textrm{s}}^{(\textrm{I})}(t)}{dt}|m\rangle,\label{eq:EoM-pm_start}
\end{align}
where $\rho_{\textrm{s}}^{(\textrm{I})}(t)$ is given by Eq. (\ref{eq:rho_s^I}). 
By substituting Eq. (\ref{eq:approx}) into Eq. (\ref{eq:EoM-pm_start}) 
and using the eigenvector of $H_{\textrm{b}}$, $|p\rangle$, 
we can express Eq. (\ref{eq:EoM-pm_start}) as follows:
\begin{align}
  &\frac{d p_{m}(t)}{dt}
  =-\int_{0}^{t}d\tau \sum\limits_{p}\notag\\
  &\times \langle m|\langle p|
  [H_{\textrm{sb}}^{(\textrm{I})}(t),
    [H_{\textrm{sb}}^{(\textrm{I})}(\tau),\rho_{\textrm{b}}\rho_{\textrm{s}}^{(\textrm{I})}(t)]]
  |p\rangle|m\rangle,\label{eq:EoM-pm}
\end{align}
where $H_{\textrm{sb}}^{(\textrm{I})}(t)$ is given by Eq. (\ref{eq:H_sb^I}). 
Combining Eq. (\ref{eq:EoM-pm}) with
Eqs. (\ref{eq:rho_s^I}) and (\ref{eq:H_sb^I}),
we obtain
\begin{align}
  \frac{d p_{m}(t)}{dt}
  =&-\int_{0}^{t}d\tau \sum\limits_{m^{\prime}}\sum\limits_{p,p^{\prime}}
  e^{i(E_{m}+E_{p}-E_{m^{\prime}}-E_{p^{\prime}})(t-\tau)}p_{p}p_{m}(t)\notag\\
  &\times 
  \langle m|\langle p|H_{\textrm{sb}}(t)|p^{\prime}\rangle|m^{\prime}\rangle
  \langle m^{\prime}|\langle p^{\prime}|H_{\textrm{sb}}(\tau)|p\rangle|m\rangle
  \notag\\
  &+\int_{0}^{t}d\tau \sum\limits_{m^{\prime}}\sum\limits_{p,p^{\prime}}
  e^{-i(E_{m}+E_{p}-E_{m^{\prime}}-E_{p^{\prime}})(t-\tau)}p_{p^{\prime}}p_{m^{\prime}}(t)\notag\\
  &\times 
  \langle m|\langle p|H_{\textrm{sb}}(\tau)|p^{\prime}\rangle|m^{\prime}\rangle
  \langle m^{\prime}|\langle p^{\prime}|H_{\textrm{sb}}(t)|p\rangle|m\rangle
  \notag\\
  &+\int_{0}^{t}d\tau \sum\limits_{m^{\prime}}\sum\limits_{p,p^{\prime}}
  e^{i(E_{m}+E_{p}-E_{m^{\prime}}-E_{p^{\prime}})(t-\tau)}p_{p^{\prime}}p_{m^{\prime}}(t)\notag\\
  &\times 
  \langle m|\langle p|H_{\textrm{sb}}(t)|p^{\prime}\rangle|m^{\prime}\rangle
  \langle m^{\prime}|\langle p^{\prime}|H_{\textrm{sb}}(\tau)|p\rangle|m\rangle
  \notag\\
  &-\int_{0}^{t}d\tau \sum\limits_{m^{\prime}}\sum\limits_{p,p^{\prime}}
  e^{-i(E_{m}+E_{p}-E_{m^{\prime}}-E_{p^{\prime}})(t-\tau)}p_{p}p_{m}(t)\notag\\
  &\times 
  \langle m|\langle p|H_{\textrm{sb}}(\tau)|p^{\prime}\rangle|m^{\prime}\rangle
  \langle m^{\prime}|\langle p^{\prime}| H_{\textrm{sb}}(t)|p\rangle|m\rangle
  ,\label{eq:EoM-pm_pre}
\end{align}
where $p_{p}=\langle p|\rho_{\textrm{b}}|p\rangle$.
Furthermore,
we can combine the first and fourth terms on the right-hand side of Eq. (\ref{eq:EoM-pm_pre})
and the second and third terms;
the results are
\begin{align}
  -\int_{0}^{t}&d\tau \sum\limits_{m^{\prime}}\sum\limits_{p,p^{\prime}}
  e^{i(E_{m}+E_{p}-E_{m^{\prime}}-E_{p^{\prime}})(t-\tau)}p_{p}p_{m}(t)\notag\\
  &\times 
  \langle m|\langle p|H_{\textrm{sb}}(t)|p^{\prime}\rangle|m^{\prime}\rangle
  \langle m^{\prime}|\langle p^{\prime}|H_{\textrm{sb}}(\tau)|p\rangle|m\rangle
  \notag\\
  -\int_{0}^{t}&d\tau \sum\limits_{m^{\prime}}\sum\limits_{p,p^{\prime}}
  e^{-i(E_{m}+E_{p}-E_{m^{\prime}}-E_{p^{\prime}})(t-\tau)}p_{p}p_{m}(t)\notag\\
  &\times 
  \langle m|\langle p|H_{\textrm{sb}}(\tau)|p^{\prime}\rangle|m^{\prime}\rangle
  \langle m^{\prime}|\langle p^{\prime}| H_{\textrm{sb}}(t)|p\rangle|m\rangle
  \notag\\
  =&-\sum\limits_{m^{\prime}}R_{mm^{\prime}}(t)p_{m}(t),
\end{align}
and
\begin{align}
  +\int_{0}^{t}&d\tau \sum\limits_{m^{\prime}}\sum\limits_{p,p^{\prime}}
  e^{-i(E_{m}+E_{p}-E_{m^{\prime}}-E_{p^{\prime}})(t-\tau)}p_{p^{\prime}}p_{m^{\prime}}(t)\notag\\
  &\times 
  \langle m|\langle p|H_{\textrm{sb}}(\tau)|p^{\prime}\rangle|m^{\prime}\rangle
  \langle m^{\prime}|\langle p^{\prime}|H_{\textrm{sb}}(t)|p\rangle|m\rangle
  \notag\\
  +\int_{0}^{t}&d\tau \sum\limits_{m^{\prime}}\sum\limits_{p,p^{\prime}}
  e^{i(E_{m}+E_{p}-E_{m^{\prime}}-E_{p^{\prime}})(t-\tau)}p_{p^{\prime}}p_{m^{\prime}}(t)\notag\\
  &\times 
  \langle m|\langle p|H_{\textrm{sb}}(t)|p^{\prime}\rangle|m^{\prime}\rangle
  \langle m^{\prime}|\langle p^{\prime}|H_{\textrm{sb}}(\tau)|p\rangle|m\rangle
  \notag\\
  =&\sum\limits_{m^{\prime}}R_{m^{\prime}m}(t)p_{m^{\prime}}(t),
\end{align}
where
\begin{align}
  R_{mm^{\prime}}(t)
  =\int_{0}^{t}d\tau \sum\limits_{p,p^{\prime}}p_{p}[
    &e^{i\Delta E(t-\tau)}
    \langle m|\langle p|H_{\textrm{sb}}(t)|p^{\prime}\rangle|m^{\prime}\rangle\notag\\
    &\times    
    \langle m^{\prime}|\langle p^{\prime}|H_{\textrm{sb}}(\tau)|p\rangle|m\rangle\notag\\
    +&e^{-i\Delta E(t-\tau)}
    \langle m|\langle p|H_{\textrm{sb}}(\tau)|p^{\prime}\rangle|m^{\prime}\rangle\notag\\
    &\times 
    \langle m^{\prime}|\langle p^{\prime}| H_{\textrm{sb}}(t)|p\rangle|m\rangle
  ],
\end{align}
and $\Delta E=E_{m}+E_{p}-E_{m^{\prime}}-E_{p^{\prime}}$.
Therefore Eq. (\ref{eq:EoM-pm_pre}) can be rewritten in the form
of Eq. (\ref{eq:Master-tdep-pre}). 

\section{Derivation of Eq. (\ref{eq:R})}

In this Appendix we derive Eq. (\ref{eq:R}).
To derive it,
we need to perform the time integration in Eq. (\ref{eq:R-pre}).
Since $H_{\textrm{sb}}(t)$ is given by Eq. (\ref{eq:Hsb}),
it is sufficient to calculate the following quantity:
\begin{align}
  I(\Delta E,\omega_{\textrm{p}})
  =\int_{0}^{t}d\tau e^{i(\Delta E+\omega_{\textrm{p}}) (t-\tau)}.\label{eq:I}
\end{align}
Indeed, by using it,
we can rewrite Eq. (\ref{eq:R-pre}) as follows:
\begin{align}
  R_{mm^{\prime}}(t)
  =&\sum\limits_{p,p^{\prime}}p_{p}
  I(\Delta E,-\omega_{\textrm{p}})
  \langle m|\langle p|H_{\textrm{sb}}^{(\textrm{abs})}|p^{\prime}\rangle|m^{\prime}\rangle\notag\\
  &\times 
  \langle m^{\prime}|\langle p^{\prime}|H_{\textrm{sb}}^{(\textrm{emi})}|p\rangle|m\rangle\notag\\
  +&\sum\limits_{p,p^{\prime}}p_{p}
  I(\Delta E,\omega_{\textrm{p}})
  \langle m|\langle p|H_{\textrm{sb}}^{(\textrm{emi})}|p^{\prime}\rangle|m^{\prime}\rangle\notag\\
  &\times
  \langle m^{\prime}|\langle p^{\prime}|H_{\textrm{sb}}^{(\textrm{abs})}|p\rangle|m\rangle\notag\\
  +&\sum\limits_{p,p^{\prime}}p_{p}
  I(-\Delta E,\omega_{\textrm{p}})
  \langle m|\langle p|H_{\textrm{sb}}^{(\textrm{abs})}|p^{\prime}\rangle|m^{\prime}\rangle\notag\\
  &\times
  \langle m^{\prime}|\langle p^{\prime}|H_{\textrm{sb}}^{(\textrm{emi})}|p\rangle|m\rangle\notag\\
  +&\sum\limits_{p,p^{\prime}}p_{p}
  I(-\Delta E,-\omega_{\textrm{p}})
  \langle m|\langle p|H_{\textrm{sb}}^{(\textrm{emi})}|p^{\prime}\rangle|m^{\prime}\rangle\notag\\
  &\times
  \langle m^{\prime}|\langle p^{\prime}|H_{\textrm{sb}}^{(\textrm{abs})}|p\rangle|m\rangle.\label{eq:R_rewrite}
\end{align}
Since Eq. (\ref{eq:I}) becomes
\begin{align}
  I(\Delta E,\omega_{\textrm{p}})
  =\frac{1}{i(\Delta E+\omega_{\textrm{p}})}[e^{i(\Delta E+\omega_{\textrm{p}})t}-1],
\end{align}
we can write Eq. (\ref{eq:R_rewrite}) as 
\begin{align}
  R_{mm^{\prime}}(t)
  =&\sum\limits_{p,p^{\prime}}p_{p}
  \frac{2\sin(\Delta E-\omega_{\textrm{p}})t}{\Delta E-\omega_{\textrm{p}}}
  \langle m|\langle p|H_{\textrm{sb}}^{(\textrm{abs})}|p^{\prime}\rangle|m^{\prime}\rangle\notag\\
  &\times
  \langle m^{\prime}|\langle p^{\prime}|H_{\textrm{sb}}^{(\textrm{emi})}|p\rangle|m\rangle\notag\\
  &+\sum\limits_{p,p^{\prime}}p_{p}
  \frac{2\sin(\Delta E+\omega_{\textrm{p}})t}{\Delta E+\omega_{\textrm{p}}}
  \langle m|\langle p|H_{\textrm{sb}}^{(\textrm{emi})}|p^{\prime}\rangle|m^{\prime}\rangle\notag\\
  &\times
  \langle m^{\prime}|\langle p^{\prime}|H_{\textrm{sb}}^{(\textrm{abs})}|p\rangle|m\rangle
  .\label{eq:R_final-pre} 
\end{align}
This is Eq. (\ref{eq:R})
because $H_{\textrm{sb}}^{(\textrm{abs})}=[H_{\textrm{sb}}^{(\textrm{emi})}]^{\dagger}$. 

\section{Derivation of an expression of the steady-state $\langle N_{\textrm{s}}\rangle$}
In this Appendix
we derive an expression of $\langle N_{\textrm{s}}\rangle=\sum_{m}N_{m}\bar{p}_{m}$.
From Eq. (\ref{eq:p_m})
we have
\begin{align}
  \langle N_{\textrm{s}}\rangle=\frac{\sum\limits_{m}N_{m}e^{-\beta (E_{m}-\mu N_{m})}}
          {\sum\limits_{m^{\prime\prime}}e^{-\beta (E_{m^{\prime\prime}}-\mu N_{m^{\prime\prime}})}}.\label{eq:N_m}
\end{align}
To perform the sums in Eq. (\ref{eq:N_m}),
we rewrite $|m\rangle$ as
$|m\rangle=|n_{1},n_{2},\cdots,n_{\infty}\rangle$,
where $n_{l}$ represents the occupation number of magnons in the state $l$.
(The description using the set $\{n_{l}\}$ may be possible
even in the presence of interactions of magnons
as long as magnons can be regarded as well-defined quasiparticles.)
As a result,
we can rewrite $N_{m}$ and $E_{m}$ as 
$N_{m}=\sum_{l}n_{l}$ and 
$E_{m}=\sum_{l}\epsilon_{l}n_{l}$, respectively, 
where $\epsilon_{l}$ represents the magnon energy in the state $l$.
By combining these equations with Eq. (\ref{eq:N_m}),
we can express the steady-state $\langle N_{\textrm{s}}\rangle$ as follows:
\begin{align}
  \langle N_{\textrm{s}}\rangle
  &=\sum\limits_{l}
  \frac{\sum\limits_{n_{l}}n_{l}e^{-\beta (\epsilon_{l}-\mu) n_{l}}}
       {\sum\limits_{n_{l}}e^{-\beta (\epsilon_{l}-\mu) n_{l}}}\notag\\
       &=\sum\limits_{l}\frac{1}{e^{\beta(\epsilon_{l}-\mu)}-1},
\end{align}
where $\mu$ is given by Eq. (\ref{eq:mu}).

\end{document}